\documentclass[12pt, preprint]{aastex}
\shorttitle{Turbulent Cooling Flows}
\shortauthors{Shadmehri & Ghanbari}
\title{COOLING FLOWS OF SELF-GRAVITATING, ROTATING,\\ VISCOUS SYSTEMS}
\author{Mohsen Shadmehri\altaffilmark{1} and Jamshid Ghanbari}
\affil{Department of Physics, School of Sciences, Ferdowsi
University, Mashhad, Iran}

\email{shad@ferdowsi.um.ac.ir\\ghanbari@ferdowsi.um.ac.ir}
\begin{document}

\begin{abstract}
We obtain self-similar solutions that describe the dynamics of a self-gravitating, rotating, viscous system. We use simplifying assumptions
but explicitly include viscosity and the cooling due to the dissipation of energy. By assuming that the turbulent dissipation of energy is as
power law of the density and the speed $v_{rms}$ and for a power-law dependence of viscosity on the density, pressure, and rotational velocity,
we investigate turbulent cooling flows. It has been shown that for the cylindrically and the spherically
cooling flows the similarity indices are the same, and they depend only on the exponents of the dissipation rate and the viscosity model.
Depending on the values of the
exponents, which the mechanisms of the dissipation and viscosity determine them, we may have solutions with different general physical properties. The conservation
of the total mass and the angular momentum of the system strongly depends on the mechanism of energy dissipation and the viscosity model.

\end{abstract}


\section{Introduction}

Detailed observations of the interstellar medium (ISM) or even the intergalactic
medium have highlighted the need to provide a description that accounts for
the turbulent pressure, thermal pressure, magnetic fields, rotation, and even stars themselves.
 A universal theory, describing the complex structures of the ISM, is far from
complete and remains a challenge for the future. However, by
dividing the ISM according to its properties, it is possible to
present satisfactory theories for particular types of ISM.
Gravitation, cooling, turbulence, and magnetic fields produce
variations in the properties of the ISM such as the density and
the temperature which in turn define the structure of the ISM.
Such structures include dense cores (e.g., Lee, Myers, \& Tafalla
1999), filamentary clouds (e.g., Schneider \& Elmegreen 1979;
Harjunpaa et al. 1999), and even disks (e.g., Padgett et al.
1999). Impressive theoretical progress has been made on the
properties and evolution of these structures during last years
under simplifying assumptions (e.g., Li 1998; Fiege \& Pudritz
2000; Tsuribe 1999). Most of the theoretical models of
interstellar clouds and clumps assume static or stationary
configurations, in which there is an equilibrium between the
self-gravity, centrifugal force and some forms of internal energy
in the cloud (e.g., Bertoldi \& McKee 1992; Chieze 1987;
Vazquez-Semadeni \& Gazol 1995; Galli et al. 2001; Tomisaka,
Ikeuchi, \& Nakamura 1988; Shadmehri \& Ghanbari 2001a).

Shu, Adams, \& Lizano (1987)
proposed a four stage scenario for the formation of an isolated low-mass star: (a) quasi-static formation and evolution
of a molecular cloud core by ambipolar diffusion; (b) dynamical collapse of the core
to a protostar and circumstellar disk; (c) breakout of a powerful bipolar
outflow; and (d) clearing of the circumstellar envelope to reveal a pre-main-sequence star.
Such theories of isolated star formation assert that gravitational collapse occurs
onto a thermally supported core and motions are quasi-hydrostatic until very late
times (e.g., Shu, Adams, \& Lizano 1987; Li 1998). Nevertheless, recent work has lead to doubts as to whether their initial condition is a reasonable starting point.
Rather than being quasi-static, the hierarchical and clumpy structures of ISM have been proposed
to result from turbulence (e.g., Falgarone, Phillips, \& Walker 1991; Elmegreen 1999).
Some authors examine the idea that ISM may not be best visualized as a system
of discrete clouds and conclude that the structures in the ISM may form as density fluctuations
induced by large-scale interstellar turbulence (e.g., Ballesteros-Paredes, Vazquez-Semadeni, \& Scalo 1999).
However, observations of infall
at small scales in isolated cores show that quasi-static course do in fact exist (Williams et al. 1999), although
it seems that for explaining large-scale motions we need an
alternative explanation (e.g., Myers \& Lazarian 1998).

Recently, we presented a non-Jeans scenario for star formation in self-gravitating
filamentary clouds (Shadmehri \& Ghanbari 2001b; hereafter SG), extending the work of  Meerson, Megged, \& Tajima (1996)
on spherical clouds. We studied quasi-hydrostatic cooling flows in these filamentary clouds.
We parameterized the cooling function as a power-law in temperature ($\Lambda\propto\rho^{2}T^{\epsilon}$), and showed that
the filament experiences radiative condensation. Furthermore, since the exponent in temperature is a free parameter, this cooling function
can also qualitatively represents turbulent energy dissipation. That is we can
use this cooling function to represent  the dissipation rate of turbulent energy by identifying
 the velocity dispersion with the temperature (see, e.g., Mac Low 1999).
 Thus, our results  also apply to a
turbulent filament, although we did not investigate this explicitly.

Myers \& Lazarian (1998) have suggested that inward, subsonic
flows arise from the local dissipation of turbulent motions in
molecular clouds. These turbulent cooling flows result from
localized dissipation of turbulence in molecular clouds, as
radiative cooling flows result from radiative cooling in clusters
of galaxies (e.g., Cowie \& Binney 1977). Myers \& Lazarian (1998)
considered only one type of turbulence (a linear superposition of
MHD waves) and only one dissipation mechanism (ion-neutral
friction). They also ignored the role of self-gravity. On the
other hand, there are competing explanations for the origin of the
clumps in the molecular clouds. A more accepted explanation is
that the clumps continuously form and disperse simply because the
gas is supersonically turbulent (e.g., Scalo 1990; Falgarone \&
Phillips 1990). Recent numerical studies of turbulence have
brought a new understanding of the physics of this complex
phenomenon. For example, recently Padoan et al. (2001) suggested
that because of the turbulent nature of supersonic motions in
molecular clouds, the dense structures such as filaments and
clumps are formed by shocks in a turbulent flow.

In this paper, we extend the work of SG to study turbulent cooling
flows in self-gravitating, rotating, viscous clouds. We adopt an
analytic approach and do not perform extensive numerical
computations. Viscosity has an important role in the behavior of
the cloud. Of course,  molecular viscosity is totally negligible
in the ISM compared to turbulent viscosity. This presents a
problem as no comprehensive treatment of turbulent viscosity
exists. For simplicity, we shall use the well-known
$\alpha-$prescription (Shakura \& Sunyaev 1973) which has been
used for modelling accretion disks and also a more general
simplified model for viscosity. This study is a first step (at
least, qualitatively) toward understanding the role of turbulent
cooling flows in self-gravitating clouds. Clearly, turbulence is a
complex phenomenon in the ISM and the results of this
investigation are {\it just}  approximations to the behavior of
real turbulent cooling flows. In general, our treatment will be
incomplete because we omit turbulent eddies with scales larger
than the size of the cloud. Thus, we consider microturbulence, and
this can only provide a first approximation to the real dynamics
of  ISM clouds. Nevertheless, our simple model provides useful
information on the importance of turbulent cooling flows in star
forming regions.

\section{General Formulation}

We shall seek similarity solutions of the set of equations describing turbulent cooling flows in
self-gravitating clouds. The simplest method of dealing with turbulence involves the assumption
 that the Navier-Stokes equation holds on each physical scale, with turbulence manifesting itself
via renormalized viscosity and heat conduction coefficients that are generally not constant.
We assume spherical (or cylindrical) symmetry and neglect heat conduction. For constructing the model, the
continuity equation, the momentum equation in the radial direction, the equation of angular
momentum conservation, the Poisson equation, and the energy equation can be re-written
explicitly in the following way:
\begin{equation}
\frac{\partial\rho}{\partial t} + \frac{1}{r^{\lambda-1}}\frac{\partial}{\partial r}(r^{\lambda-1} \rho v_{r})=0,
\end{equation}
\begin{equation}
\frac{\partial v_{r}}{\partial t} + v_{r}\frac{\partial v_{r}}{\partial r} + \frac{1}{\rho}\frac{\partial p}{\partial r} +
\frac{\partial \Psi}{\partial r} = \frac{v_{\varphi}^{2}}{r},
\end{equation}
\begin{equation}
\frac{\partial (r v_{\varphi})}{\partial t} + v_{r} \frac{\partial (r v_{\varphi})}{\partial r} = \frac{1}{r^{\lambda - 1} \rho}
\frac{\partial}{\partial r}(\nu \rho r^{\lambda + 1} \frac{\partial \Omega}{\partial r}),
\end{equation}
\begin{equation}
\frac{1}{r^{\lambda-1}}\frac{\partial}{\partial r}(r^{\lambda-1}\frac{\partial\Psi}{\partial r}) = 4\pi G\rho,
\end{equation}
\begin{equation}
\frac{1}{\gamma-1}(\frac{\partial p}{\partial t} + v_{r}\frac{\partial p}{\partial r}) + \frac{\gamma}{\gamma-1}\frac{p}{r^{\lambda-1}}
\frac{\partial}{\partial r}(r^{\lambda-1}v_{r}) + \Lambda(\rho, T) = 0,
\end{equation}
where $\rho$, $v_{r}$, $v_{\varphi}$, $p$, and $\Psi$ denote the gas density, radial velocity, rotational velocity, pressure, and
gravitational potential, respectively. Also, $\Omega = \frac{v_{\varphi}}{r}$ is the angular velocity and $\lambda$ determines
the dimensionality of the model, with $\lambda$ being $2$ in cylindrical geometry and $3$ in spherical geometry.

To close this system of equations, the form of  and viscosity $(\nu)$ cooling function $(\Lambda)$ must be known. As was mentioned
above, our understanding of turbulent viscosity is incomplete, and for this reason we adopt an empirical prescription.
For example, the $\alpha -$prescription
assumes that the {\it molecular} viscosity can be replaced by an isotropic $turbulent$ viscosity, and so
\begin{equation}
\nu = \alpha \frac{p}{\rho \Omega},
\end{equation}
where $\alpha$ is a constant (Shakura \& Sunyaev 1973). It has been shown that $\alpha$ is not in general  constant, but rather
depends on a number of factors (see, e.g., Brandenburg 1998). However,  for simplicity, we will assume that the parameter $\alpha$ is
constant. Also, we consider a more general viscosity model than $\alpha -$model, viz.,
\begin{equation}
\nu = \alpha p^{\tau} \rho^{\beta} \Omega^{\eta},
\end{equation}
where $\tau$, $\beta$, and $\eta$ are three arbitrary exponents. A viscosity model similar to this form has been used by Begelman \& Meier (1982).
By imposing additional requirements like conservation of the total mass or angular momentum of the system, we shall find relations between
these three exponents. It is clear that the $\alpha-$model corresponds  to
\begin{equation}
\tau = 1, \beta = -1, \eta = -1,
\end{equation}
and is one member of this family. Note that the constant $\alpha$ in equation (7) is in general different from the $\alpha$ in equation (6);
they are the same only when the exponents have values given by
 equation (8). We use $\alpha$ as a free parameter to study the effect of viscosity.

For dissipation, we could use the Mac Low (1999) result locally, which would give a dissipation time that would be always comparable to the local
crossing time. In reality, the dissipation rate is not just a function of temperature. The $T$ varies all over because of shocks, and most of the dissipation is in shocks
, not in a uniform thermal medium. So, if we have some characteristic scale, like the radius around a filament or the distance from
the center of a sheet, we would take $\rho v_{rms}^{3}/L$ as the dissipation rate for speed $v_{rms}$. Then $L = r$ of where we consider
in the filament, for example. Also, some authors have found power-law relations between $v_{rms}$ and $L$ by doing numerical simulations (e.g.,
Vazquez-Semadeni \& Avila-Reese 2001).
We note that it is possible by knowing the $v_{rms}$, define the temperature $T$. Neglecting any viscous and external heating, however, it is possible to fit the cooling function (dissipation rate) by a power law,
\begin{equation}
\Lambda(\rho, T)= A_{\epsilon}\rho^{2} T^{\epsilon},
\end{equation}
where $A_{\epsilon}$ and $\epsilon$ are constants. This general
form enable us to study the behavior of the system by changing the
values of $\epsilon$ and also $A_{\epsilon}$. In fact, the
mechanism of turbulent dissipation determines the values of these
parameters.

In certain optically thin systems, we can also use  equation (9)  to approximate the radiative cooling function.
Radiative cooling almost always involves a two body process, and as such also depends of the square of the density. The exponent
of temperature $\epsilon$ depends on the regime under consideration.  In molecular clouds, cooling is dominated by
dust or CO line emission, and the range of $\epsilon$ is from $1.5$ to $3$ (Goldsmith \& Langer 1978).
Also, for hot plasmas of cosmic composition, we have $\epsilon \backsimeq \frac{1}{2}$ for $T > 10^{7} K$ when free-free emission dominates, and
$\epsilon \backsimeq -\frac{1}{2}$ for $10^{5} K < T < 10^{7} K$ when line cooling dominates (Gaets \$ Salpeter 1983). This approach is useful
for qualitative investigation of the effects of radiative cooling, but we must be aware of its limitations.

We are interested in quasi-hydrostatic flows in the clouds. For quasi-hydrostatic flows, equation (2) becomes
\begin{equation}
\frac{1}{\rho}\frac{\partial p}{\partial r} + \frac{\partial \Psi}{\partial r} = \frac{v_{\varphi}^{2}}{r}.
\end{equation}
To simplify the equations, we make the following substitutions:
\begin{eqnarray}
\rho \rightarrow \hat{\rho} \rho, p \rightarrow \hat{p}p, \Psi \rightarrow \hat{\Psi}\Psi,
T \rightarrow \hat{T} T,  v_{r,\varphi} \rightarrow \hat{v} v_{r,\varphi}, r \rightarrow \hat{r} r, t \rightarrow \hat{t} t,
\end{eqnarray}
where

\begin{equation}
\hat{\rho} = \rho_{0}, \hat{p} = p_{0}, \hat{T} = T_{0}, \hat{\Psi} = \frac{p_{0}}{\rho_{0}},
\hat{r} = (\frac{p_{0}}{4\pi G\rho_{0}^{2}})^{\frac{1}{2}},
\hat{t} = \frac{p_{0}}{(\gamma - 1)A_{\epsilon}\rho_{0}^{2}T_{0}^{\epsilon}}, \hat{v} = \frac{\hat{r}}{\hat{t}}, m = \frac{\hat{v}^{2} \rho_{0}}{p_{0}}.
\end{equation}
Under these transformations, equations (1) and (10) do not change, but the other equations become
\begin{equation}
\frac{\partial (r v_{\varphi})}{\partial t} + v_{r} \frac{\partial (r v_{\varphi})}{\partial r} = \frac{\alpha}{m r^{\lambda - 1} \rho}
\frac{\partial}{\partial r}( p^{\tau} \rho^{\beta +1} \Omega^{\eta} r^{\lambda + 1} \frac{\partial \Omega}{\partial r}),
\end{equation}
\begin{equation}
\frac{1}{r^{\lambda-1}}\frac{\partial}{\partial r}(r^{\lambda-1}\frac{\partial\Psi}{\partial r}) = \rho,
\end{equation}
\begin{equation}
\\\frac{\partial p}{\partial t} + v_{r}\frac{\partial p}{\partial r} + \gamma \frac{p}{r^{\lambda-1}}
\frac{\partial}{\partial r}(r^{\lambda-1}v_{r}) + \rho^{2-\epsilon} p^{\epsilon} = 0.
\end{equation}
Notice the equations are not invariant under the transformation $t \rightarrow -t$, $v_{r} \rightarrow -v_{r}$, and $v_{\varphi} \rightarrow -v_{\varphi}$, so the
solutions can not be time-reversible. Thus, these equations describe a nonlinear evolving system.

\section{Similarity Solutions}
\subsection{Analysis}

In following the  nonlinear evolution of dynamically evolving systems, the technique of self-similar analysis is useful, as it allows a set of
partial differential equations, such as those above, to be  transformed into a set of ordinary differential equations.
A similarity solution, although constituting only a limited part of the problem, is often useful in understanding  the basic behaviour of the system.
In order to seek similarity solutions to the above equations, we introduce a similarity variable $\xi$ as
\begin{equation}
\xi = \frac {r}{(t_{0} - t)^{n}},
\end{equation}
and assume that each physical  quantity is given by the following form:
\begin{equation}
\rho(r, t) = (t_{0} - t)^{\nu_{1}} R(\xi),
\end{equation}
\begin{equation}
p(r, t) = (t_{0} - t)^{\nu_{2}} P(\xi),
\end{equation}
\begin{equation}
v_{r}(r, t) = (t_{0} - t)^{\nu_{3}} V(\xi),
\end{equation}
\begin{equation}
v_{\varphi}(r, t) = (t_{0} - t)^{\nu_{4}} \Phi(\xi),
\end{equation}
\begin{equation}
\Psi(r, t) = (t_{0} - t)^{\nu_{5}} S(\xi),
\end{equation}
where the exponents $n$, $\nu_{1}$, $\nu_{2}$, $\nu_{3}$, $\nu_{4}$, and $\nu_{5}$ are constants which must be determined and $t<t_{0}$. By substituting the above equations into
equations (1), (10), (13), (14), and (15), we obtain the general results:
\begin{equation}
\nu_{1} = \frac{2n(1-\epsilon) -1}{\epsilon }, \nu_{2} = \frac{2n(2 - \epsilon) - 2}{\epsilon},
\nu_{3} = n - 1, \nu_{4} = \frac{2n - 1}{2\epsilon}, \nu_{5} = \frac{2n - 1}{\epsilon},
\end{equation}
and,
\begin{equation}
\\n = \frac{\epsilon - (\beta + 2\tau + \frac{\eta}{2})}{(\eta + 2\beta + 2\tau + 2)\epsilon - (\eta + 2\beta + 4\tau)}.
\end{equation}
In the specific case of the $\alpha-$model, we have $\tau = 1, \beta = \eta = -1$, which gives:
\begin{equation}
\\n = \frac{2\epsilon - 1}{2(\epsilon -1)}, \nu_{1} = -2, \nu_{2} = \frac{3 - 2\epsilon}{\epsilon -1}, \nu_{3} = \frac{1}{2(\epsilon - 1)}, \nu_{4} =
\frac{1}{2(\epsilon - 1)}, \nu_{5} = \frac{1}{\epsilon - 1}.
\end{equation}
It is interesting that these similarity indices are independent of the dimensionality of the system, i.e. $\lambda$,
but depend strongly on the exponents of the viscosity and the turbulent cooling function. Thus, in both cylindrical and spherical {\it just} the mechanisms
of viscosity and dissipation of energy are important.

We can write the equations for the dependence of the
physical quantities on the similarity variable as:
\begin{equation}
\\ -\nu_{1} R + n \xi \frac{dR}{d\xi} + \frac{1}{\xi^{\lambda - 1}} \frac{d}{d\xi}(\xi^{\lambda - 1} RV) = 0,
\end{equation}
\begin{equation}
\frac{1}{R}\frac{dP}{d\xi} + \frac{dS}{d\xi} = \frac{\Phi^{2}}{\xi},
\end{equation}
\begin{equation}
\\ -\nu_{4} \xi \Phi + n \xi^{2} \frac{d\Phi}{d\xi} + V \frac{d}{d\xi}(\xi \Phi) = \frac{\alpha}{m} \frac{1}{\xi^{\lambda - 1} R}
\frac{d}{d\xi}(\xi^{\lambda - \eta + 1} P^{\tau} R^{\beta + 1} \Phi^{\eta}\frac{d}{d\xi}(\frac{\Phi}{\xi})),
\end{equation}
\begin{equation}
\\ -\nu_{2} P + n \xi \frac{dP}{d\xi} + V \frac{dP}{d\xi} + \gamma \frac{P}{\xi^{\lambda - 1}} \frac{d}{d\xi}(\xi^{\lambda - 1} V) +
R^{2 - \epsilon} P^{\epsilon} = 0,
\end{equation}
\begin{equation}
\frac{1}{\xi^{\lambda - 1}}\frac{d}{d\xi}(\xi^{\lambda - 1}\frac{dS}{d\xi}) = R.
\end{equation}

This is a system of non-linear ordinary differential equations which has some critical points. Since the similarity indices are known, it is
possible to study the general physical properties of the solutions.
We see that the solution for each physical quantity retains a similar form as the flow evolves, but the characteristic length-scale
of the flow increases proportionally to $(t_{0} - t)^{n}$, where equation (23) gives the value of $n$.
As we require that $n$ be finite and nonzero, this relation shows that $\epsilon$
may have any value except $\epsilon = \frac{\eta + 2\beta + 4\tau}{\eta + 2\beta + 2\tau + 2}$ or $\epsilon = \beta + 2\tau + \frac{\eta}{2}$.
 The sign of $n$ determines whether the cloud expands or collapses; if
 $n$ is positive the radius of the cloud $r_{c}$ decreases with time and the central density $\rho_{c}$ increases with time.
In the $\alpha-$model, $n$ depends only on  $\epsilon$, and is negative for $\epsilon$ between $\frac{1}{2}$ and $1$,
 and positive otherwise. Also, in this case, the density at the center of the cloud increases in proportion to $(t_{0} - t)^{-2}$, irrespective of
the value of $\epsilon$.

We now consider boundary conditions, such as conservation of the total mass, angular momentum and
constant external pressure. Conservation of mass is an important physical constraint, and, as we shall show,
fortuitously simplifies the equations and
allows us to obtain some analytical results. The integrals representing the total mass ($M$) and angular momentum ($J$) are the following:
\begin{equation}
\\M = \int \rho r^{\lambda - 1} dr,
\end{equation}
\begin{equation}
\\J = \int \rho \Omega r^{\lambda + 1} dr.
\end{equation}
Thus, the total mass $M$ (nondimensional) is proportional to
$(t_{0} - t)^{\nu_{1} + \lambda n}$, and the total angular
momentum $J$ (nondimensional) proportion to $(t_{0} - t)^{\nu_{1}
+ \nu_{4} + (\lambda + 1)n}$. As these relations show the
conservation of $M$ or $J$ depends on the dimensionality of the
system and the exponents  of viscosity and cooling rate. First, we
study the properties of the solutions for the $\alpha-$model.
There is no value of $\epsilon$ for which the mass of a
cylindrical cloud to be conserved and for spherical clouds the
mass of the cloud is conserved only for $\epsilon = -\frac{1}{2}$.
The total angular momentum is conserved with $\epsilon =
\frac{\lambda - 4}{2(\lambda - 1)}$. However, when the total mass
of a spherical cloud is conserved ($\epsilon = -\frac{1}{2}$), the
total angular momentum decreases: $J\varpropto (t_{0} -
t)^{\frac{1}{3}}$. Another interesting case is a cylindrical cloud
with general form of viscosity, in which the constraint of
constant total mass gives
\begin{equation}
\eta + 2\beta + 2\tau = 0.
\end{equation}
Note that this constraint involves only the indices of the
viscosity model and  not the index of the dissipation model,
$\epsilon$. Thus, this constraint is independent  of the mechanism
of dissipation. Under this constraint, equations (22) give the
similarity indices:
\begin{equation}
\\n = \frac{1}{2}, \nu_{1} = -1, \nu_{2} = -1, \nu_{3} = -\frac{1}{2}, \nu_{4} = 0, \nu_{5} = 0.
\end{equation}
Also, in this case, the total angular momentum decreases: $J \varpropto (t_{0} - t)^{\frac{1}{2}}$.

Constant external pressure is another
interesting case in which the cloud is surrounded by an ambient gas: $p(r \rightarrow \infty) =$ const. Of course, the total mass need not be
conserved, as inflow or outflow of the gas is allowed.
In the $\alpha-$model this corresponds to $\epsilon =  \frac{3}{2}$, regardless of dimensionality, and
 the total mass and the angular momentum of the system decreases:
 $M\varpropto (t_{0} -t)^{2(\lambda -1)}$ and $J\varpropto (t_{0} - t)^{(2\lambda + 1)}$.

From the above, it is clear that the solutions show different
behaviors depending on the values of the exponents of the
dissipation of energy and the viscosity (that is, on the
dominating mechanisms of  dissipation and  viscosity). Also, these
results are independent of the magnitude of the turbulent
viscosity.

\subsection{Numerical solutions}
Once $\epsilon$, $\eta$, $\beta$, and $\tau$ are selected, we can solve the set of ordinary differential equations (25)-(29).
The full range of possibilities becomes enormous if we regard  all the exponents of the viscosity model and the cooling
rate as free parameters. For this reason, we shall investigate only the case
of constant total mass. First, we study cylindrical cooling flows with
constant total mass, in which the constraint given by equation (32) must be satisfied. If
we require a finite density and zero velocity at the cloud center, equation (25) gives
\begin{equation}
\\V = - n \xi,
\end{equation}
where $n = \frac{1}{2}$. Substituting this equation into equation (28), gives an algebraic relation between the similarity
functions of the pressure and the density:
\begin{equation}
\\P = P_{0} R^{q},
\end{equation}
where $P_{0} = (\gamma - 1)^{\frac{1}{\epsilon - 1}} $ and $q = \frac{\epsilon - 2}{\epsilon - 1}$. It is interesting that this  relation
 for turbulent and rotating cooling flows is the same as the one derived by SG for radiative cooling flows.
Note that there is no critical value for $\gamma$, because for all values of $\gamma > 1$, the relation between the density and the pressure is well-defined.
Also, $\epsilon$ may have any values except $\epsilon = 1$ and $2$. Going back to equations (27) and (29) and using this result, we have:
\begin{equation}
\frac{d}{d\xi}(\xi^{3 - \eta} R^{q\tau + \beta + 1} \Phi^{\eta} \frac{d}{d\xi}(\frac{\Phi}{\xi})) + \frac{m}{2\alpha P_{0}^{\tau}} \xi^{2} \Phi R = 0,
\end{equation}
\begin{equation}
\frac{d}{d\xi}(\Phi^{2} - q P_{0} \xi R^{q-2} \frac{dR}{d\xi}) - \xi R = 0.
\end{equation}

We have to solve these equations subject to the boundary conditions:
\begin{equation}
\\R(\xi = 0) = 1, (\frac{dR}{d\xi})_{\xi = 0} = 0, \Phi(\xi = 0) = 0, (\frac{d\Phi}{d\xi})_{\xi = 0} = 0.
\end{equation}
Of course, there is another constraint: conservation of mass.
Since we are not interested in the exact value of the mass of the
system, we don't parameterize the solutions by the total mass. In
fact, we integrate the equations by the Runge-Kutta method to seek
solutions satisfying the boundary conditions and then we can
calculate the mass of the system. However, it could be simply
checked that if $q = \frac{4 - \tau}{\tau - 2}$, then the above
equations are invariant under a homology transformation. Indeed,
if $R(\xi)$ and $\Phi(\xi)$ are a set of solutions of the
equations, then $A^{\frac{1}{q-2}}R(A \xi)$ and $A^{\frac{2q -
3}{2(q-2)}}\Phi(A \xi)$ are also a set of solutions, where $A$ is
an arbitrary constant. Using this transformation, it is possible
to find $A$ so that the solutions are satisfying the boundary
conditions and the constraint of constant mass. Thus, if $\tau$
and $q$ are satisfying the relation $q=\frac{4-\tau}{\tau-2}$, we
can parameterize the solutions by the line mass of the cylindrical
cloud.

In Figure 1, we represent the distributions of the density
$R(\xi)$ in the self-similar space for $\gamma=\frac{5}{3}$,
$m=4.0$, $\alpha=0.007$, $\tau=\frac{1}{2}$, $\eta=-1$, $\beta=0$
and different values of $\epsilon=3$, $4$, $-3$, $-4$. They are
obtained by solving the equations with the fourth-order
Runge-Kutta method. This Figure shows that the general behaviors
of the similarity density function $R(\xi)$ for different values
of $\epsilon$ are almost the same. In fact, all the solutions tend
to zero as $\xi \rightarrow \infty$. It means that we can define
the radius of the cloud by the condition $R(\xi_{c}) \backsimeq
0$, where $r_{c}=\xi_{c} (t_{0} - t)^{\frac{1}{2}}$ (see equation
(33)).

Figures 2 and 3 show the behavior of the similarity rotational
velocity function $\Phi(\xi)$ for the parameters the same as
Figure 1. From Figure 1 and 2 we see that whole of the cloud is
rotating for $\epsilon = 3$ and $4$. But Figure 3 shows that only
part of the cloud is rotating for $\epsilon = -3$ and $-4$. Also,
each profile shows a maximum value for the rotational velocity.

For the $\alpha-$model viscosity, we showed that only for
$\epsilon=-\frac{1}{2}$ the spherically cooling flows describe a
system with constant total mass and this value of $\epsilon$
corresponds to free-free emission for cosmic compositions. Figure
4 shows the density profile of spherically symmetric flows for
different values of $\alpha$. As the Figure shows the profile of
$R(\xi)$ hardly depends on the value of $\alpha$. It means that
the distribution of the mass in spherically symmetric cooling
flows is independent of the magnitude of viscosity. But as Figure
5 shows, the distribution of angular momentum in such systems
strongly depends on the magnitude of viscosity. In Figure 5, we
can see the normalized rotational velocity $\Phi(\xi)$ for
different values of $\alpha$ and it shows as the magnitude of
viscosity increases, the maximum value of the profile of
$\Phi(\xi)$ increases.

\section{Discussion and Conclusions}

We have explored in this paper a new set of similarity solutions of the equations relevant to cooling flows of self-gravitating, rotating, viscous
systems. Analogously to the case with no rotation and viscosity (SG; Meerson, Megged, \& Tajima 1996), we were able to obtain similarity
solutions which describe cooling flows in rotating and viscous systems. However, we did not investigate all of the solutions of the system. In fact, we
restrict ourselves to the quasi-hydrostatic flows with constant total mass under a wide range of the viscosity and the cooling models. When
the central density is high and the velocities small, we can apply our solutions which are regular at the center. Even if we solve the equations
without the quasi-steady assumption by including the time dependent terms in the momentum equation, we shall find the same similarity indices as equation (22).

It may seems that physical constraints on our solutions, such as
mass or angular momentum conservation, limit the acceptable range
of similarity indices. For example, the $\alpha-$viscosity law is
incompatible with mass conservation in our models. But it does not
mean that the self-similarity assumption is not a good assumption.
As for the viscosity model, we mentioned that since within the
turbulent viscosity assumption the main difficulty is to point out
a plausible source of turbulence, empirical viscosity
prescriptions (e.g., $\alpha-$model) have been broadly used.
Although the $\alpha-$model enables a reasonable global
description of stationary, thin and non self-gravitating disks
(Frank et al., 1992), we must note that the $\alpha-$model is only
an {\it empirical} prescription. Few improvements  of the
$\alpha-$prescription have been made (Narayan 1992; Narayan et al,
1994; Godon 1995) and it has been shown that in general $\alpha$
is not constant (Brandenburg 1998). Thus, just according to
considerations about viscosity models, we can not investigate the
validity of our solutions. Due to these facts, we introduced a
general form for viscosity law, i.e. equation (7). However, one
should demonstrate that the acceptable viscosity laws have
reasonable dependencies on physical quantities. But such study is
out of the scope of this paper.

Recently, Elmegreen (1999) investigated the role of wave-driven turbulence in interstellar clouds. In his study, a
dense region forms because of a convergence and high pressure from external magnetic waves and the thermal cooling that follows
at the compressed interface. However, turbulent dissipation at the interface is not as important as thermal cooling
for this density enhancement (Elmegreen 1999). Thus, if in our model we consider the dissipation rate simply as thermal cooling function
(not as turbulent dissipation rate), probably it will be possible to apply the results of this study for the compressed interfaces in
Elmegreen's model.

In fact, some cosmological structures, diffuse HI mediums, and
molecular clouds, all of which show some cooling properties. Thus,
our solutions might have applications to the formation of these
structures at early stages. In this regard, this work is an
extension of the same cooling flow problem considered in SG and
Meerson, Megged, \& Tajima (1996) in order to study the effects of
viscosity and rotation on cooling flows. For filamentary clouds,
SG showed that if the mass of the system is conserved, then the
density at the center increases ($\rho_{c} \varpropto (t_{0} -
t)^{-1}$) and the radius of the filament decreases ($r_{c}
\varpropto (t_{0} - t)^{\frac{1}{2}}$), irrespective of the
$\epsilon$, that is, the mechanism of cooling. We showed that if
we consider the rotation and viscosity, conservation of the total
mass of the system depends on the mechanism of cooling and
viscosity model. For example for the $\alpha-$ model, all the
cooling mechanisms lead to a system with nonconstant mass. But for
the other models of viscosity which the exponents satisfy into
equation (32), the total mass of the system is conserved and we
have: $\rho_{c} \varpropto (t_{0} - t)^{-2}$ and $r_{c} \varpropto
(t_{0} - t)^{\frac{1}{2}}$. Also, for spherical cooling flow the
value of $\epsilon=-\frac{1}{2}$ corresponds to constant mass. In
this case, we showed that $\rho_{c} \varpropto (t_{0} - t)^{-2}$
and $r_{c} \varpropto (t_{0} - t)^{\frac{2}{3}}$. However,
Meerson, Megged, \& Tajima (1996) showed that for nonviscous and
nonrotating spherical cooling flows, the behaviors of $r_{c}$ and
$\rho_{c}$ depend on $\epsilon$ and the mass of system is
conserved: $r_{c} \varpropto (t_{0} - t)^{\frac{1}{\epsilon + 2}}$
and $\rho_{c} \varpropto (t_{0} - t)^{-\frac{3}{2 + \epsilon}}$.

A dispersion relation has been derived for gravitational instabilities in a medium with cloud collisional cooling by Elmegreen (1989). The
cooling function due to the cloud-cloud collisions with an isotropic, Maxwellian distribution of cloud velocities is a power-law of the density
and the temperature (Elmegreen 1987), similar to the form of our cooling rate in this paper. Thus, in a cloudy medium without much star formation
activity (i.e., without heating), the energy dissipation is given by equation (9). Elmegreen showed that the regions with such conditions
will clump into cloud complexes on a variety of scales. Since on the largest scales, the rotation and shear of the Galaxy become important,
our results describe the general properties of structure formation in such regions.

This simple model has nevertheless some limitations. We have not included explicit heating terms in our energy equation, in order to minimize
the number of parameters. However, we can consider additional terms in the governing equations. For example, one could invoke the flow inertia
in the momentum equation, or heat conduction in the energy equation. Although the similarity indices would be selected in these cases, we can consider
only limited values of the exponents of the cooling function and the viscosity model. Also, the assumption of similarity solutions, could be dropped, although we should
not forget the complex nature of the governing partial differential equations. Also, we assumed a power law for the form of cooling rate and viscosity and the
results depend strongly on the form of these functions. However, as we mentioned in SG, it seems that we can use piece-wise solutions. It means that
for each part of the evolution of the solutions, we can find the suitable exponents and then join one set of solutions to the another one. Thus, it will be possible
to consider a much wider range of cooling functions and viscosity models with the same analytical-type of solution.
\acknowledgments{This work was completed while one of the authors (MS) was visiting Instituto de Astronomia, UNAM, and he
acknowledges gratefully the hospitality of Professor Enrique Vazquez-Semadeni during his visit. We also thank Alan Watson and Bruce Elmegreen
for carefully reading early versions of the manuscript and providing us with detailed comments.
MS acknowledges research studentship from Ferdowsi University of Mashhad.}

\begin{figure}
\plotone{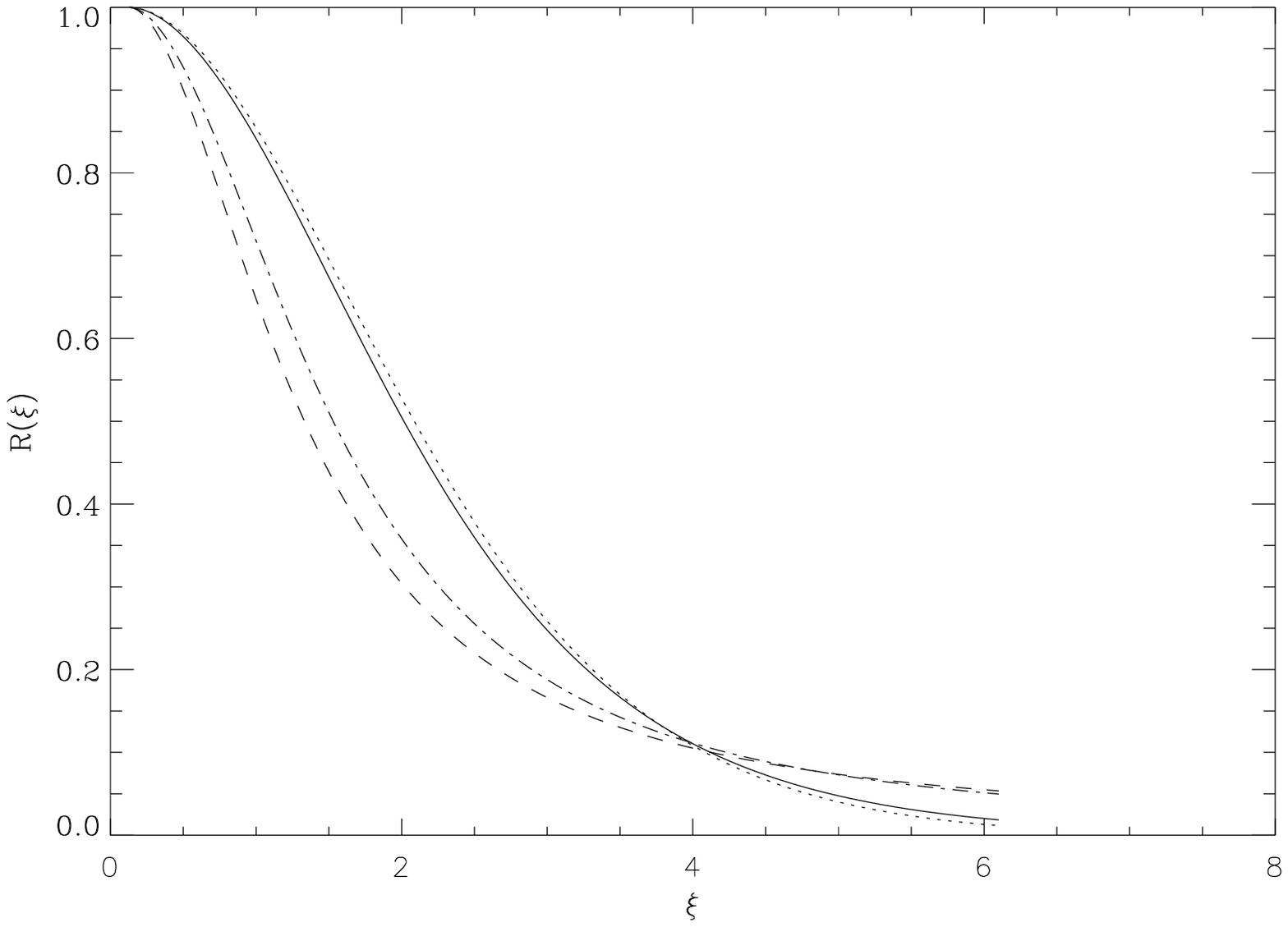}
\caption{Similarity solutions for the cylindrically symmetric quasi-hydrostatic flows for $\gamma=\frac{5}{3}$, $m=4.0$, $\alpha=0.007$, $\tau=\frac{1}{2}$, $\eta = -1$, and
 $\beta = 0$. Shown are the normalized density $R(\xi)$ for different values of $\epsilon=-4$ ({\it solid curve}), $-3$ ({\it dotted curve}), $4$ ({\it dashed curve }), and $3$ ({\it long-dashed curve}). All
 the solutions tend to zero as $ \xi \rightarrow \infty $.}
\end{figure}
\begin{figure}
\plotone{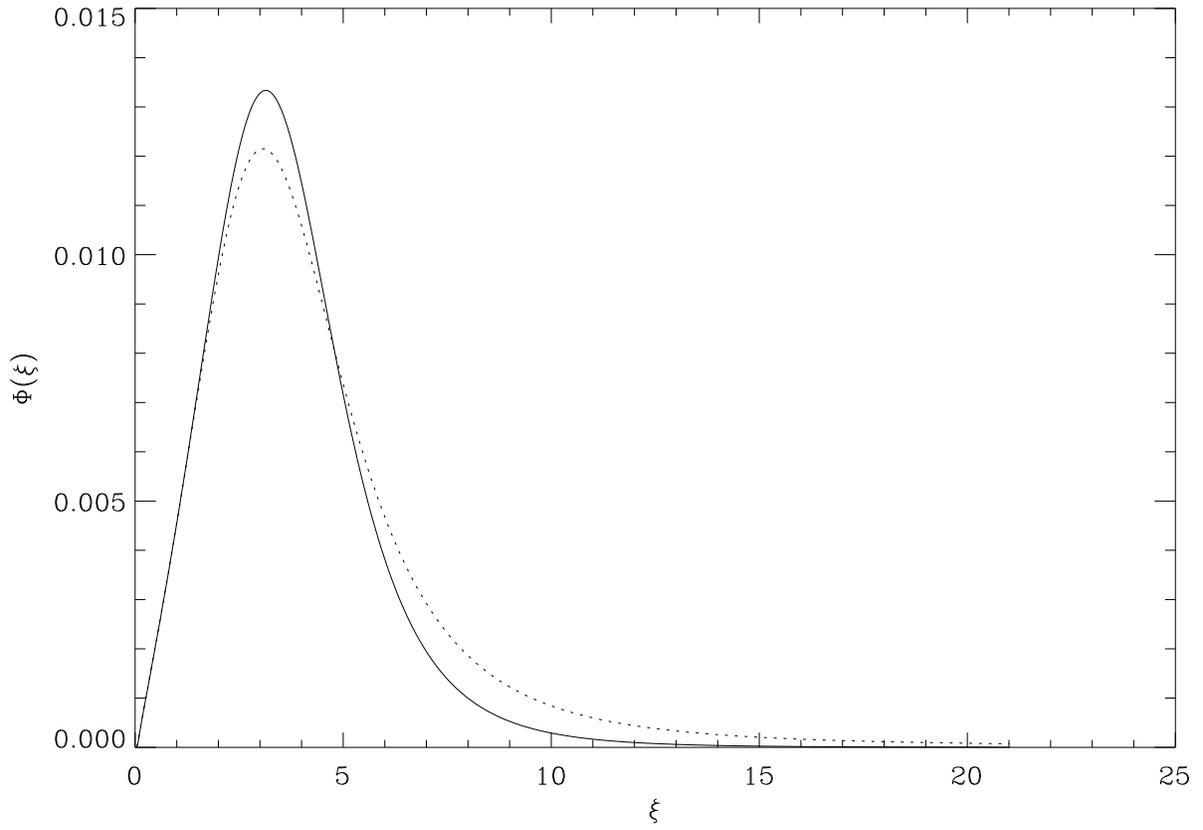}
\caption{Similarity solutions for the cylindrically symmetric quasi-hydrostatic flows for $\gamma=\frac{5}{3}$, $m=4.0$, $\alpha=0.007$, $\tau=\frac{1}{2}$, $\eta = -1$, and
 $\beta = 0$. Shown are the normalized rotational velocity $\Phi(\xi)$ for different values of $\epsilon=4$ ({\it solid curve}), $3$ ({\it dotted curve}).}
\end{figure}
\begin{figure}
\plotone{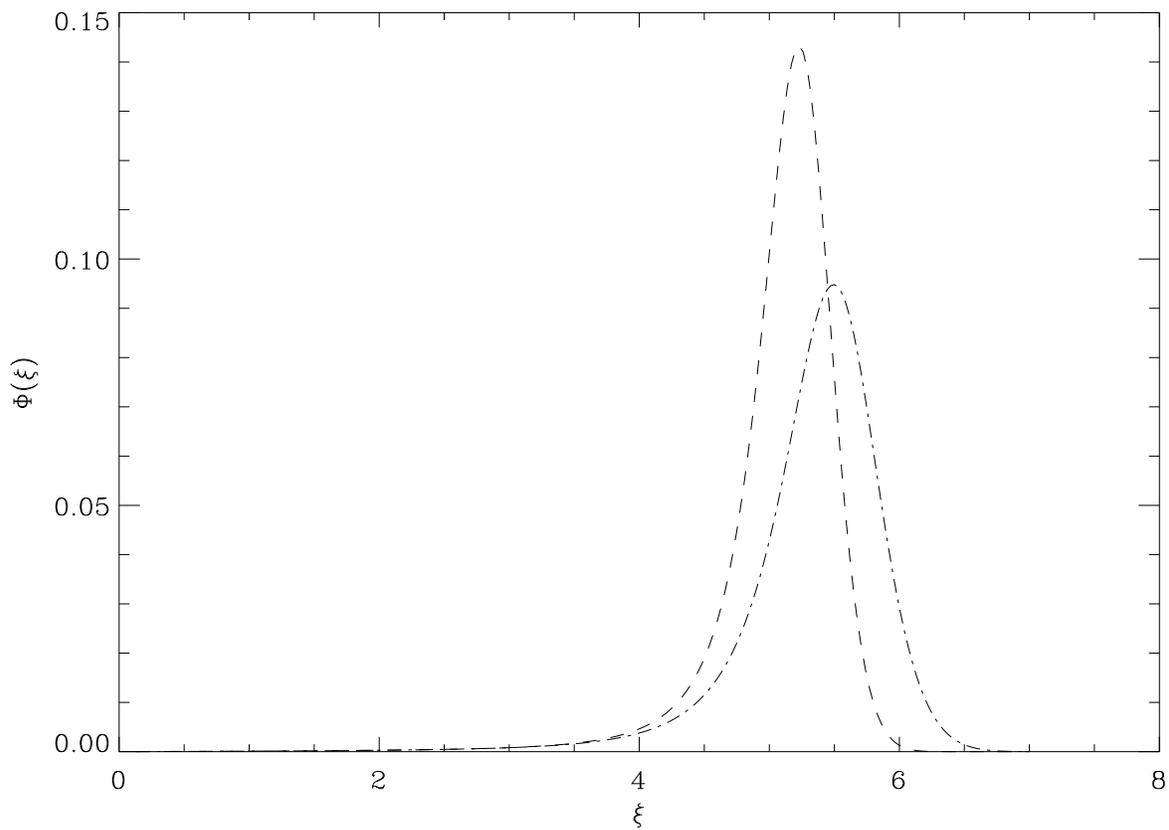}
\caption{Same as Figure 2, but for different values of $\epsilon = -3$ ({\it dashed}) and $-4$ ({\it long-dashed}).}
\end{figure}
\begin{figure}
\plotone{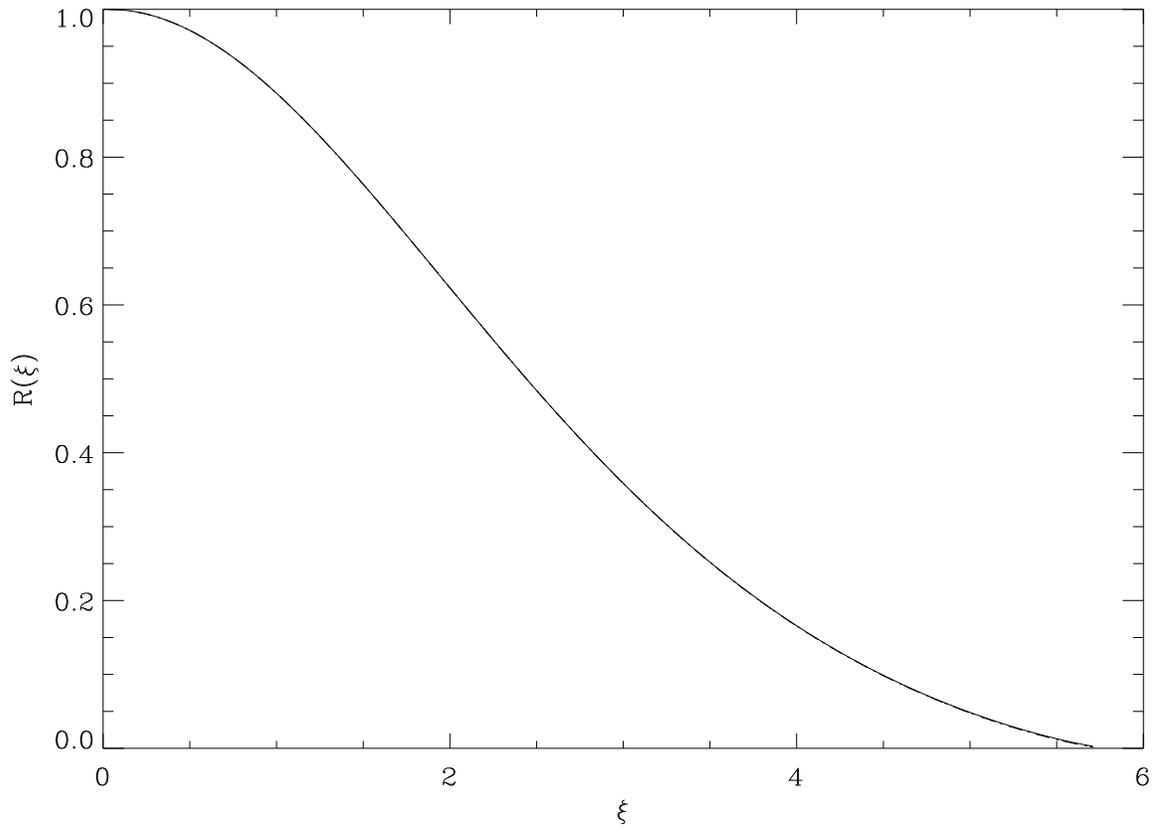}
\caption{Similarity solutions for the spherically symmetric flows for $\gamma=\frac{5}{3}$, $m = 4.0$, $\epsilon=-\frac{1}{2}$, and $\alpha-$ model viscosity.
Shown are the normalized density $R(\xi)$ for different values of $\alpha = 0.005$ ({\it solid curve}), $0.05$ ({\it dotted curve}), and
$0.1$ ({\it dashed curve}). We can see that $R(\xi)$ hardly depends on $\alpha$.}
\end{figure}
\begin{figure}
\plotone{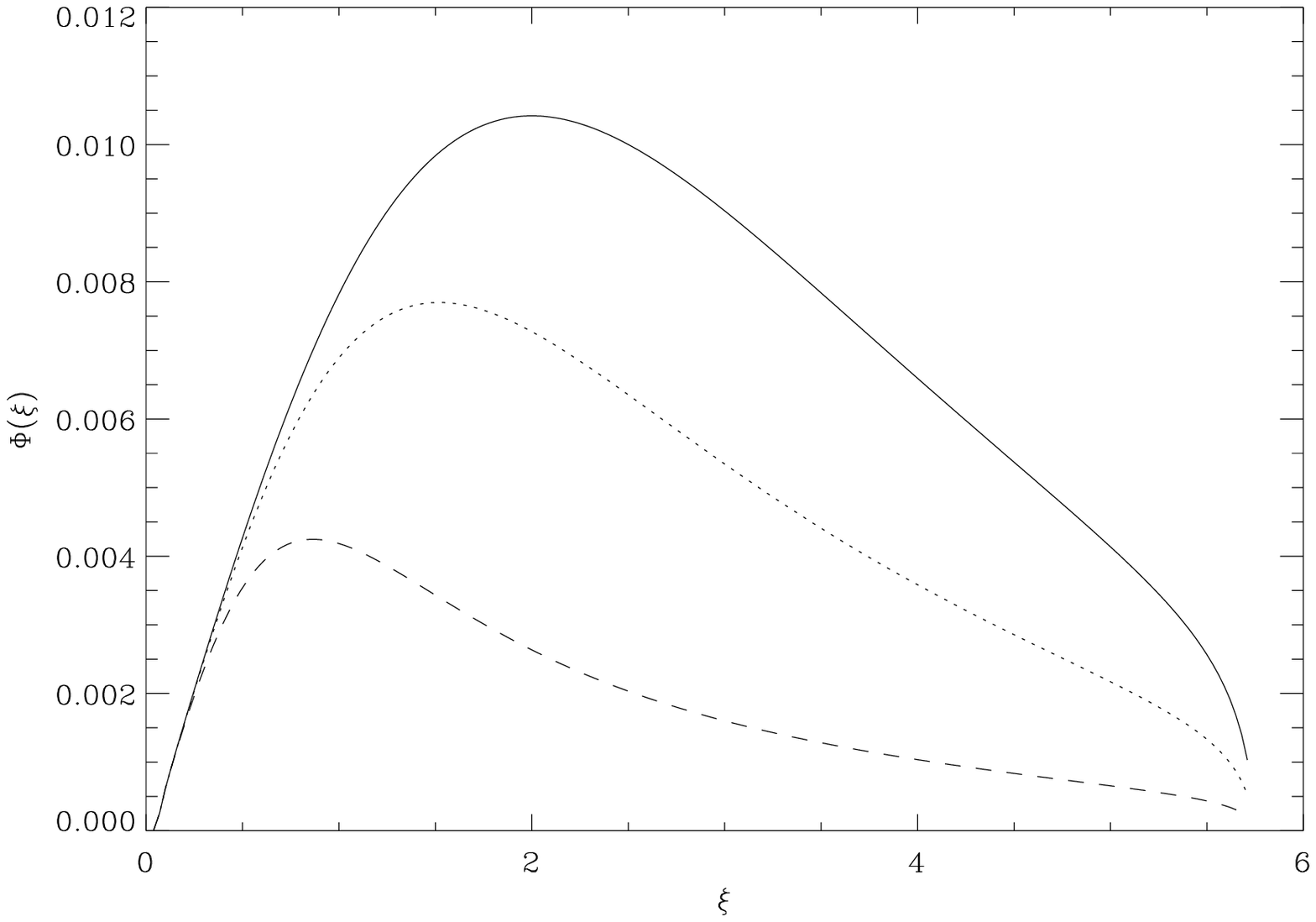}
\caption{Similarity solutions for the spherically symmetric flows for $\gamma=\frac{5}{3}$, $m = 4.0$, $\epsilon=-\frac{1}{2}$, and $\alpha-$ model viscosity.
Shown are the normalized rotational velocity $\Phi(\xi)$ for different values of $\alpha = 0.005$ ({\it solid curve}), $0.003$ ({\it dotted curve}), and
$0.001$ ({\it dashed curve}).}
\end{figure}

\begin{references}
\reference{b1} Avila-Reese, V., \& Vazquez-Semadeni, E. 2001, ApJ,
553, 645

\reference{b2} Ballesteros-Paredes, J., Vazquez-Semadeni, E., \&
Scalo,  J. 1999, ApJ, 515, 286

\reference{b3} Begelman, M.C., \& Meier, D.L. 1982, ApJ, 253, 873

\reference{b4} Bertoldi, F., \& McKee, C.F. 1992, ApJ, 395, 140

\reference{b5} Brandenburg, A., 1998, in: Abramowicz M.A.,
Bjornsson G., Pringle J.E., eds.,'Theory of Black Hole Accretion
Discs', p.61

\reference{b6} Chieze, J.P. 1987, A\&A, 171, 225

\reference{b7} Cowie, L.L., \& Binney, J. 1977, ApJ, 215, 723

\reference{b8} Elmegreen, B.G. 1987, ApJ, 312, 626

\reference{b9} Elmegreen, B.G. 1989, ApJ, 344, 306

\reference{b10} Elmegreen, B.G. 1999, ApJ, 527, 266

\reference{b11} Falgarone, E., \& Phillips, T.G. 1990, ApJ, 359,
344

\reference{b12} Falgarone, E., Phillips, T., \& Walker, C.K. 1991,
ApJ, 378, 186

\reference{b13} Fiege, J.D., \& Pudritz, R.E. 2000, MNRAS, 311,
105

\reference{b14} Frank, J., King, A., \& Raine, D. 1992, Accretion
Power in Astrophysics (Cambridge: Cambridge Univ. Press)

\reference{b15} Gaetz, T.J., \& Salpeter, E.E. 1983, ApJS, 52, 155

\reference{b16} Galli, D., Shu, F.H., Laughlin, G., \& Lizano, S.
2001, ApJ, 551, 367

\reference{b17} Godon, P. 1995, MNRAS, 277, 157

\reference{b18} Goldsmith, P.F., \& Langer, W.D. 1978, ApJ, 222,
881

\reference{b19} Harjunpaa, P., Kass, A.A., Carlqvist, P., \& Gahm,
G.F. 1999, A\&A, 349, 912

\reference{b20} Lee, C.W., Myers, P.C., \& Tafalla, M. 1999, ApJ,
526, 788

\reference{b21} Li, Z.-Y 1998, ApJ, 497, 850

\reference{b22} Mac Low, M.M. 1999, ApJ, 524, 169

\reference{b23} Meerson, B., Megged, E., \& Tajima T. 1996, ApJ,
457, 321

\reference{b24} Myers, P.C., \& Lazarian, A. 1998, ApJ, 507, L157

\reference{b25} Narayan, R. 1992, ApJ, 394, 261

\reference{b26} Narayan, R., Loeb, A., \& Kumar, P. 1994, ApJ,
431, 359

\reference{b27} Neufeld, D.A., \& Kaufman, M.J. 1993, ApJ, 418,
263

\reference{b28} Padgett, D.L., Brandner, W., Stapelfeldt, K.R.,
Strom, S.E., Terebey, S., \& Koerner, D. 1999, AJ, 117, 1490

\reference{b29} Padoan, P., Juvela, M., Goodman, A.A., \&
Nordlund, A. 2001, ApJ, 553, 227

\reference{b30} Scalo, J.M. 1990, in Physical Processes in
Fragmentation and Star Formation, ed. R. Capuzzo-Dolcetta, C.
Chiosi, \& A. Di Fazio (Dordrecht: Kluwer), 151

\reference{b31} Schneider, S., \& Elmegreen, B. 1979, ApJS, 41, 87

\reference{b32} Shadmehri, M., \& Ghanbari, J. 2001a, ApJ, 557,
1028

\reference{b33} Shadmehri, M., \& Ghanbari, J. 2001b, A\&SS, 278,
347 (SG)

\reference{b34} Shakura, N.I., \& Sunyaev, R.A. 1973, A\&A, 24,
337

\reference{b35} Shu, F.H., Adams, F.C., \& Lizano, S. 1987,
ARA\&A, 25, 23

\reference{36} Spitzer, L. 1978, Physical Processes in the
Interstellar Medium, Wiley, New York

\reference{b37} Tomisaka, K., Ikeuchi, S., \& Nakamura, T. 1988,
ApJ, 326, 208

\reference{b38} Tsuribe, T. 1999, ApJ, 527, 102

\reference{b39} Vazquez-Semadeni, E., \& Gazol, A. 1995, A\&A,
303, 204

\reference{b40} Williams, J.P., Myers, P.C., Wilner, D.J., \& Di
Francesco, J. 1999, ApJ, 513, L61
\end{references}
\end{document}